\documentclass[envcountsame]{llncs}
\usepackage[letterpaper,hmargin=1in,vmargin=1.25in]{geometry}
\usepackage{amsmath,amssymb,bm}

\usepackage{algorithm}
\usepackage{algpseudocode}
\usepackage{url,ifthen}
\usepackage[T1]{fontenc}
\usepackage{color}
\usepackage{amsmath,amssymb,mathrsfs}
\usepackage{bm}
\usepackage{graphicx}

\newcommand{\wt}{\mathsf{w}}

\newcommand{\code}[1]{\ensuremath{\mathcal{#1}}}

\newcommand{\eqdef}{\mbox{~$\stackrel{\scriptstyle{\rm def}}{=}$}~}

\newcommand{\FF}[1]{\ensuremath{\mathbb{F}_#1}}

\newboolean{anonymous}
\setboolean{anonymous}{false}

\begin{document}

\pagestyle{plain}

\title{\bf Post-Quantum Cryptography: Code-based Signatures}

\author{
    Pierre-Louis Cayrel
\and
    Mohammed Meziani
}
\institute{
    CASED -- Center for Advanced Security Research Darmstadt\\
    Mornewegstrasse, 64293 Darmstadt, Germany\\
    \url{pierre-louis.cayrel@cased.de}\\
    \url{mohammed.meziani@cased.de}
}

\date{\today}

\maketitle

\begin{abstract}
$\ $\\ This survey provides a comparative overview of code-based signature schemes with respect to security and performance. Furthermore,
we explicitly describe serveral code-based signature schemes with additional properties such as identity-based, threshold ring and blind signatures.
\end{abstract}

\section{Introduction}\label{sec:intro}

Secure digital signature are essential components of IT-security solutions, and several schemes, such as the Digital Signature Algorithm DSA and the Elliptic Curve Digital Signature Algorithm ECDSA are already used in practice. The security of such schemes relies on the hardness of the discrete logarithm problem, either in the multiplicative group of a prime field, or in a subgroup of points of an elliptic curve over a finite field. These computational assumptions, however, could be broken in a quantum setting by Shor's algorithm \cite{shor97}, which was proposed in 1997. Moreover, this algorithm succeeds in polynomial time. Therefore, new, quantum-attack-resistant signature schemes must be designed. Code-based cryptosystems are promising alternatives to classical public key cryptography, and they are believed to be secure against quantum attacks. Their security is based on the conjectured intractability of problems in coding theory, such as the syndrome decoding problem, which has been proven to be NP-complete by Berlekamp, McEliece, and Van Tilborg \cite{berlekamp-mceliece-vantilborg}.

In 1978, McEliece \cite{McEliece78} first proposed an asymmetric cryptosystem based on the coding theory, which derives its security from the general decoding problem. The general idea is to first select a particular (linear) code for which an efficient decoding algorithm is known, and then to use a trapdoor function to disguise the code as a general linear code. Though numerous computationally-intensive attacks against the scheme appear in the literature \cite{BLP08,finiasz-sendrier}, no efficient attack has been found to date.

The McEliece encryption scheme is not invertible, and therefore it cannot be used for authentication or for signature schemes; this is indeed why very few signature schemes based on coding theory have been proposed. This problem remained open until 2001, when Courtois et al. \cite{courtois-finiasz-sendrier} showed how to achieve a code-based signature scheme whose security is based on the syndrome decoding problem. While this problem is NP-complete, constructions based on it are still inefficient for large numbers of errors.

A few code-based signature schemes with additional properties, most of them based on the construction of \cite{courtois-finiasz-sendrier}, have recently been published. Lattice-based digital signature schemes for a post-quantum age are described in \cite{buchmann-lindner-ruckert-schneider}. This paper describes code-based solutions.

\vspace{-0.3cm}

\subsubsection{Contribution and Organisation:}$\ $\\
After recalling some basic definitions and notations in Section \ref{sec:prelim}, we discuss the various code-based signature schemes, starting with CFS, Stern, and KKS in Section \ref{sec:codesign}. 
 In Section \ref{sec:codebasedwithSP}, we describe \textit{all} code-based signature schemes with additional properties, and we conclude in Section \ref{sec:conclusion}.

\newpage
%#########  Preminaliries ############
\section{Coding theory background }\label{sec:prelim}

This section recalls some basic definitions and then lists some instances of hard problems in coding theory.

\begin{definition}\textbf{(Linear Code)}
An $(n,k)$-code over $\mathbb{F}_q$ is a linear subspace $\mathcal{C}$ of the linear space $\mathbb{F}^n_q$. Elements of $\mathbb{F}^n_q$ are called \emph{words}, and elements of $\mathcal{C}$ are \emph{codewords}. We call $n$ the \emph{length}, and $k$ the \emph{dimension} of $\mathcal{C}$.
\end{definition}
\begin{definition}\textbf{(Hamming distance, weigth)}
The \emph{Hamming distance} $d(\mathsf{x},\mathsf{y})$ between two words $\mathsf{x}, \mathsf{y}$ is the number of positions in which $\mathsf{x}$ and $\mathsf{y}$ differ. That is, $d(\mathsf{x}, \mathsf{y}) = |\{i \; : \; x_i \neq y_i\}|$, where $\mathsf{x} = (x_1, \dots , x_n)$ and $\mathsf{y} = (y_1, \dots , y_n)$. Here, we use $|S|$ to denote the number of elements, or cardinality, of a set $S$. In particular, $d(\mathsf{x},\mathbf{0})$ is called the Hamming weigth of $\mathsf{x}$, where $\mathbf{0}$ is the vector containing $n$ $0$'s. The \emph{minimum distance} of a linear code $\mathcal{C}$ is the minimum Hamming distance between any two distinct codewords. 
\end{definition}
\begin{definition}\textbf{(Generator matrix)}
 A \emph{generator matrix} of an $(n,k)$-linear code $\mathcal{C}$ is a $k \times n$ matrix $\mathsf{G}$ whose rows form a basis for the vector subspace $\mathcal{C}$. We call a code \emph{systematic} if it can be characterized by a generator matrix $\mathcal{C}$ of the form $\mathsf{G} = (\mathsf{I_{k \times k}} | \mathsf{A_{k \times (n-k)}})$, where $\mathsf{I_{k \times k}}$ is the $k \times k$ identity matrix and $\mathsf{A}$, an $k \times (n-k)$ matrix.
\end{definition}

\begin{definition}\textbf{(Parity-check matrix)}
A \emph{parity-check} matrix of an $(n,k)$-linear code $\mathcal{C}$ is an $(n-k) \times n$ matrix $\mathsf{H}$ whose rows form a
basis of the orthogonal complement of the vector subspace $\mathcal{C}$, i.e. it holds
that, $\mathcal{C} = \{\mathsf{c} \in \mathbb{F}_q^n \; : \; \mathsf{Hc^T} = \mathbf{0}\}$.
\end{definition}

In what follows, we recall several NP-complete problems in coding theory. Note that NP-completeness ensures the impossibility to solve a problem in polynomial time \emph{in the worse case}. In other words, the property ensures the existence of \emph{some} hard instances, not the hardness of \emph{every} instance.

\begin{definition}{\textbf{(Binary Syndrome Decoding (SD) problem)}}
\begin{itemize}
\item \textbf{Input:} An $r\times n$ matrix $\mathsf{H}$ over $\mathbb{F}_2$, a target binary vector $\mathsf{s} \in \mathbb{F}_2^r$, and an integer $t > 0$.
\item \textbf{Question:} Is there a binary word $\mathsf{x}\in\mathbb{F}_2^n$ of weight $\leq t$, such that $\mathsf{s = Hx^T}$ ?
\end{itemize}
\end{definition}
This problem has been proved to be NP-Complete by Berlekamp, McEliece, and van Tilborg \cite{berlekamp-mceliece-vantilborg}.
In 1994, Barg \cite{Barg94} extended this result of Berlekamp, McEliece, and van Tilborg over $\mathbb{F}_q$ by proving that the following problem, called $q$-ary Syndrome Decoding ($q$-SD) problem, is NP-complete.
\begin{definition}\textbf{($q$-ary Syndrome Decoding ($q$-SD) problem)}
\begin{itemize}
\item \textbf{Input:} An $r\times n$ matrix $\mathsf{H}$ over $\mathbb{F}_q$, a target vector $\mathsf{s} \in \mathbb{F}_q^r$, and an integer $t > 0$.
\item \textbf{Question:} Is there a word $\mathsf{x}\in\mathbb{F}_q^n$ of weight $\leq t$, such that $\mathsf{s = Hx^T}$ ?
\end{itemize}
\end{definition}
To end this section, we state the Goppa Code Distinguishing (GD) problem~:% which has been proved NP-complete in \cite{Fin04}.
\begin{definition}\textbf{(Goppa Code Distinguishing (GD) problem)}
\begin{itemize}
\item \textbf{Input:} An $(n-k)\times n$ binary matrix $\mathsf{H}$.
\item \textbf{Question:} Is $\mathsf{H}$ a parity check matrix of a $(n,k)$-Goppa code or of a random $(n,k)$-code ?
\end{itemize}
\end{definition}

\newpage
\section{Code based signature schemes}\label{sec:codesign}
During the last twenty years several (linear)-code-based signature schemes were proposed; the first attempts were due to Xinmei Wang \cite{Wan90}, followed by Harn and Wang \cite{HW92} and Alabbadi and Wicker \cite{AW92a}. Unfortunately, the security of these constructions cannot be reduced to the hardness of the problems above, and the schemes were proved insecure \cite{Wan90,HW92}.

Several signature schemes based on these problems were subsequently designed; we outline these below.

\subsection{Courtois et. al's scheme}

Unlike RSA, one of the major obstacles to the widespread use of the McEliece or the Niederreiter cryptosystems was the one-to-one nature of the encryption algorithms, i.e, a random word $x\in\mathbb{F}_2^n$ that is encrypted to, say, $y$ is not necessary decodable. That is, the Hamming distance between $y$ and any codeword is greater than the error capability of the code. This is due the fact that the cardinality of decodable words is very small. To fix this problem, Courtois, Finiasz, and Sendrier \cite{courtois-finiasz-sendrier} (CFS) suggested a method, named \emph{complete decoding}, which increases the correction capability in order to find the nearest word to a given codeword with high probability.

The CFS signature scheme uses Goppa codes that are subfield subcodes of particular alternant code \cite{macwilliams-sloane}. For given integers $m$ and $t$, binary Goppa codes are of length $n = 2^m$ , of dimension $k = n - mt$, and are $t$-correcting. The basic idea of the CFS signature scheme is to find parameters $n$, $k$, and $t$ such that the Niederreiter scheme described in Algorithm \ref{nieder} is practically invertible. 

\begin{algorithm}
\caption{The Niederreiter PKC}\label{nieder}
\begin{algorithmic}
\State  \textbf{Key Generation:}
\State \quad - Consider an $(n,k)$-code $\mathcal{C}$ over $\mathbb{F}_q$ having a decoding algorithm $\gamma$.
\State \quad - Construct an $(n-k)\times n $ parity check matrix $\widetilde{H}$ of $\mathcal{C}$.
\State \quad - Choose randomly an $(n-k)\times(n-k)$ invertible matrix $Q$ over $\mathbb{F}_q$.
\State \quad - Choose randomly an $n\times n$ permutation matrix $P$ over $\mathbb{F}_q$.
\State \quad \quad- \textbf{The public key:} $H = Q\widetilde{H}P$
\State \quad \quad- \textbf{The private key:} $(P,\widetilde{H},Q,\gamma)$
\State  \textbf{Encryption:} To encrypt a message $x\in \mathbb{F}_{q}^{n}$ of weight $t$
\State \quad - Compute $y=Hx^{T}$.
\State  \textbf{Decryption:} To decrypt a cipher $y\in \mathbb{F}_q^{n-k}$ s.t. $y=Hx^{T}$
\State \quad - Compute $Q^{-1}y\ (=\widetilde{H}Px^{T})$
\State \quad - Find $Px^{T}$ from $Q^{-1}y$ by applying $\gamma$
\State \quad - Find $x$ by applying $P^{-1}$ to $Px^{T}.$
\end{algorithmic}
\end{algorithm}

A CFS signature on a message $M$ -- see algorithm \ref{CFSAlgo} -- is generated by hashing $M$ to a syndrome and then trying to decode it. However, for a $t$-error correcting Goppa code of length $n=2^m$, only about $1/t!$ of the syndromes are decodable. Thus, a counter is appended to $M$, and the signer updates the counter until the hash value is decodable. The signature consists of both the syndrome's weight $t$ error pattern and the counter value.

\begin{algorithm}
\caption{The CFS signature }\label{CFSAlgo}
\begin{algorithmic}
\State  \textbf{Key Generation:}
\State  - Pick random parity check matrix $\widetilde{H}$ of $(n, k)$-binary,
\State  \quad $t$ error-correcting Goppa code with decoding algorithm $\gamma$.
\State  - Construct binary matrices $Q$, $H$ and $P$ as in Algorithm \ref{nieder}.
\State  \textbf{Signature:} To sign a message $M$
\State  (1) $i \leftarrow i+1$
\State  (2) $x^\prime = \gamma\left(Q^{-1}h(h(m)\|i)\right)$
\State  (3) if no $x^\prime$ was found go to 1
\State  - Output $(i, x^\prime P)$
\State  \textbf{Verification:}
\State  - Compute $s^\prime = H{x^\prime}^T$ and $s = h(h(m) \| i)$.
\State  - The signature is valid if $s $ and $s^\prime$ are equals.
\end{algorithmic}
\end{algorithm}
\textbf{Security.} The authors of \cite{finiasz-sendrier} show an attack against the CFS scheme due to Daniel Bleichenbacher. This attack is based on an 'unbalanced' Generalized Birthday Attack. Therefore, the values of $m$ and $t$ used by CFS have been changed. For a security of more than $2^{80}$ binary operations, \cite{finiasz-sendrier} proposed new parameters of: $m = 21$ and $t = 10$; $m = 19$ and $t = 11$; or $m = 15$ and $t = 12$. Furthermore, the authors of modified CFS (mCFS) \cite{dallot} give a security proof in the random oracle model, where the counter is randomly chosen in $\{1,\ldots,2^{n-k}\}$.

\subsection{Stern's identification scheme}
In 1993, Stern \cite{stern} presented a 3-pass zero-knowledge protocol which is closely related to the Niederreiter cryptosystem. This protocol aims at enabling a {\em prover} $P$ to identify himself to a {\em verifier} $V$. Its principle is as follows: Let $H$ be an $(n-k)\times n$ binary matrix common to all users, where $n$ and $k$ are integers s.t. $k\leq n$. Each prover $P$ has an $n$-bit secret key $\textsf{s}$ of weight $t$ and an $(n-k)$-bit {\em public identifier} $\textsf{y}$ satisfying $\textsf{y}={H}\textsf{s}^{T}$. When $P$ needs to authenticate to $V$ as the owner of $\textsf{y}$, then $P$ and $V$ run the Algorithm \ref{SternId}. It was shown in \cite{stern} that the probability that an adversary successfully impersonates an honest prover is $2/3$.

\begin{algorithm}
\caption{Stern's Scheme}\label{SternId}
\begin{algorithmic}
\State \textbf{Key Generation :} Given binary random $(k,n)$-code with parity-check matrix $H$, secure hash function $h$.
\State \quad - \textbf{Private key:} $s \in \mathbb{F}^n_2$, such that $\wt(s) = t$
\State \quad - \textbf{Public key:} $y \in \mathbb{F}^{n-k}_2$, such that $Hs^T=y$
\State \textbf{Commitments:}
\State \quad - $P$ chooses randomly $u \text{ from } \mathbb{F}_2^n$ and $\sigma$ permutation over $\left\{ 1, \ldots, n\right\}$
\State \quad - $P$ computes the commitments $c_1$, $c_2$, and $c_3$ as follows:
\State \quad \quad \quad $c_1 = h(\left( \sigma, Hu^T \right))$, $c_2 = h(\sigma(u))$, $c_3 = h(\sigma \left( u \oplus s\right))$
\State \quad - $P$ sends $c_1$, $c_2$, and $c_3$ to $V$
\State \textbf{Challenge:} $V$ randomly chooses $b\in\{0,1,2\}$ and sends it to $P$
\State \textbf{Response:}
\State \quad - If $b=0$: $P$ sends $u$ and $\sigma$ to $V$
\State \quad - If $b=1$: $P$ sends $u \oplus s$ and $\sigma$ to $V$
\State \quad - If $b=2$: $P$ sends $\sigma(u)$ and $\sigma(s)$ to $V$
\State \textbf{Verification :}
\State \quad - If $b=0$: $V$ checks if $c_1$ and $c_2$ were honestly computed
\State \quad - If $b=1$: $V$ checks if $c_1$ and $c_3$ were honestly computed
\State \quad - If $b=2$: $V$ checks if $c_2$ and $c_3$ were honestly computed and $\wt(\sigma(s)) =t$
\end{algorithmic}
\end{algorithm}

In 1995, V\'eron \cite{veron} proposed a dual version of Stern's scheme, which, unlike other schemes based on the SD problem, uses a generator matrix of a random binary linear code. This allows, among other things, for an improved transmission rate.

It is possible to convert Stern's construction into a signature algorithm using the Fiat-Shamir method \cite{FiaSham86}: the verifier-queries are replaced by values suitably derived from the commitments and the message to be signed. In this case, however, the signature is large, of roughly 120 Kbits.

A variation of the Stern construction using double circulant codes is proposed in \cite{gaborit-girault}. The circulant structure of the public parity-check matrix allows for an easy generation of the whole binary matrix with very little memory storage. They propose a scheme with a public key of 347 bits and a private key of 694 bits. We can also imagine a construction based on quasi-dyadic codes as proposed in \cite{misoczki-barreto}.

A secure implementation \cite{cayrel-gaborit-prouff} of Stern's scheme uses quasi-circulant codes. This scheme also inherits Stern's natural resistance to leakage attacks such as SPA and DPA.

\subsection{Kabatianskii et al.'s scheme}

Kabatianskii, Krouk, and Smeets (KKS) \cite{kks97} proposed a signature scheme based on arbitrary linear error-correcting codes. Actually, they proposed three versions (using different linear codes) presented in the sequel and all have one point in common: the signature is a codeword of a linear code. We give a full description of the KKS scheme which is illustrated in Algorithm \ref{KKSAlgo}.

First consider a code $\code{C}$ defined by a random parity-check matrix $H$; let $d$ be a good estimate of its minimum distance. Next, consider a linear code $\code{U}$ of length $n'\le n$ and dimension $k$ defined by a generator matrix $G = [g_{i,j}]$. We suppose that there exist integers $t_1$ and $t_2$ s.t. $t_1 \le \wt{(u)} \le t_2$ for any non-zero codeword $u \in \code{U}$.

Let $J$ be a subset of $\{1,\ldots{},n\}$ of cardinality $n'$, $H(J)$ be the sub matrix of $H$ consisting of the columns $h_i$ where $i \in J$, and define an $r \times n'$ matrix $F \eqdef H(J) G^T$. Define a $k \times n$ matrix $G^* = [g^*_{i,j}]$ with $g^*_{i,j} =  g_{i,j}$ if $ j \in J$ and $g^*_{i,j} = 0$ otherwise. The KKS-signature is $\sigma$ $= mG^*$ for any $m \in \FF{q}^k$. The main difference with Niederreiter signature occurs in the verification step where the receiver checks that: $t_1 \le \wt{(\sigma)} \le t_2 ~\mbox{ and }~ F \cdot{} m^T = H \cdot{} \sigma^T.$

\begin{algorithm}
\caption{The KKS Signature }\label{KKSAlgo}
\begin{algorithmic}
\State  \textbf{Key Generation:}
\State  - Pick random $(n,n-r)$ code $\mathcal{C}$, then choose secretly and randomly:
\State  \quad (1) Generator matrix $G$ of an $(n',k)$ code $\code{U}$ with $n'< n$ and such that $\forall v\in \mathcal{V}, v\neq 0 \quad t_1\leq \wt(v)\leq t_2$
\State  \quad (2) Subset $J$ of $\{1,\cdots ,n\}$ of cardinality $n'$
\State  - Form the submatrix $H(J)$ consisting of the columns $h_i$ of a parity check matrix $H$ of $\mathcal{C}$ where $i \in J$
\State  - Define the matrix $F$ as $F = H(J)G^T$.
\State  \quad \textbf{Private key:} $(J, G)$
\State  \quad \textbf{Public Key:} $(F, H, t_1, t_2 )$
\State  \textbf{Signature:} To sign a message $m$
\State  \quad  (1) Calculate $\sigma^* = m\cdot G$
\State  \quad  (2) Produce $\sigma$ such that
 \[ \sigma_i = \left\{
\begin{array}{ll}
\sigma^*_i & \text{if}\quad i\in J\\
0 & \text{if}\quad j\notin J
\end{array}
\right. \]
\State  \textbf{Verification:} Given $(\sigma,m)$ test whether the following holds:
\State  \quad (1) $H\cdot\sigma^T = F\cdot m$
\State  \quad (2) $t_1 \le \wt{(\sigma)} \le t_2$
\end{algorithmic}
\end{algorithm}
\textbf{Security.} The authors of \cite{kks97} proposed four KKS-signature schemes: KKS-1, KKS-2, KKS-3, KKS-4, which are claimed to be as secure as the Niederreiter scheme if the public parameters do not provide any information. Unfortunately, in \cite{cayrel-otmani-vergnaud} the author showed that a generated KKS-signature discloses a lot of information about the secret set $J$, and so an adversary can find the secret matrix $G$ with a very high probability. Indeed, an attacker needs about $2^{77}$ binary operations and at most 20 signatures to break the original KKS-3 scheme. For this reason, the authors of \cite{cayrel-otmani-vergnaud} suggest new parameters for a security of 40 signatures, as follows: $n = 2000$, $k = 160$, $n' = 1000$, $r = 1100$, $t_1 = 90$ and $t_2 = 110$.

\section{Code based Signature schemes with additional properties}\label{sec:codebasedwithSP}

There exist just a few code-based signature schemes with special properties (SP) up to date, namely blind, (threshold) ring signatures, and identity-based signature schemes. By comparison, classical cryptography includes more than sixty classes of signature schemes, some with special properties such as group- or proxy signature. This variety reflects the wide range of application scenarios.

In recent years, existing signature schemes were combined with specific protocols in order to achieve enhanced code-based constructions with additional features, such as anonymity. The properties of the underlying basic scheme could be e.g. authentication and non-repudiation. In what follows, we give a state of the art of such signature schemes.

\subsection{Ring Signatures}

The concept of ring signatures was firstly introduced in 2001 by Rivest, Rivest, and Tauman \cite{RST2001}. Such signature schemes allow signers of a document to remain anonymous in a group of users, called a \emph{ring}. As opposed to group signatures, no group manager, group setup procedure, cooperation, and revocation mechanisms are needed in ring signatures: the signer specifies an arbitrary ring and then signs on its behalf without permission or assistance from other users. To generate a valid signature, users need their private keys and some other members' public keys.

\subsubsection{Zheng et al.'s scheme} In \cite{ZhengLC07}, Zheng, Li, and Chen (ZLC) proposed the first code-based ring signature, which extends the CFS signature scheme and is based on the syndrome decoding problem. To describe the ZLC signature, we use the following notations. Let $N$ and $l$ be the number of potential signers and of signers participating in the signature-generating, respectively. Denote by $S_i$ and $S_r$ a potential signer and the ring signer, respectively. Let $M$ be a message and $h$, a hash function of range $\FF{2}^{n-k}$. Write the concatenation of $s_1$ and $s_2$ as $(s_1|s_2)$; let $u\xleftarrow[]{R} \mathcal{U}$ indicate that $u$ is randomly selected from a set $\mathcal{U}$. The ring signer and all other potential signers run Algorithm \ref{RingZLC} to generate a ring signature on $M$.

\begin{algorithm}

\footnotesize{
\caption{The ZLC ring Signature }\label{RingZLC}
\begin{algorithmic}
\State  \textbf{Key Generation:} Potential signers $S_i$ generate their private/public keys as in the CFS algorithm (Alg.\ref{CFSAlgo}):
\State \quad \quad- \textbf{The public key:} $H_i = Q_i\widetilde{H_i}P_i$
\State \quad \quad- \textbf{The private key:} $(P_i,\widetilde{H_i},Q_i,\gamma_i)$
\State \textbf{Signature:} To sign message $M$
\State \quad \quad (1) \textit{Initialization:} For $j = 0,1,2,\cdots$
\State \quad \quad \quad \quad - $\bar x_j \xleftarrow[]{R} \left\{0,1\right\}^{n-k}$
\State \quad \quad \quad \quad - Set $ x_{r+1,j} = h(N|h(M)|\bar x_j)$
\State \quad \quad (2) \textit{Generating ring sequences:} For $j = 0,1,2,\cdots$
\State \quad \quad \quad \quad - $z_{i,j} \xleftarrow[]{R} \left\{0,1\right\}^n$ s.t. $\wt(z_{i,j}) = t$
\State \quad \quad \quad \quad - Set $ x_{i+1,j} = h \left(N|h(M)|H_i\cdot z^T_{i,j} \oplus x_{i,j}\right)$
\State \quad \quad (3) Find an $j_0$ s.t. $ x_{r,j_0} \oplus \bar x_{j_0}$ is decodable
\State \quad \quad (4) Apply the decoding algorithm to get an $z_{r,j_0}$ s.t. $H_r\cdot z^T_{r,j_0} = x_{r,j_0} \oplus \bar x_{j_0}$
\State \quad \quad (5) Compute the index $I_{z_{i,j_0}}$ corresponding to $z_{i,j_0}$
\State \quad \quad (6) The ring signatutre: $(x_{0,j_0},I_{z_{1,j_0}},\cdots,I_{z_{l-1,j_0}})$
\State \textbf{Verification:} Given $(x_{0,j_0},I_{z_{1,j_0}},\cdots,I_{z_{(l-1),j_0}})$
\State \quad \quad (1) Derive $z_{i,j_0}$ from $I_{z_{i,j_0}}$ for each $i\in\{0,1,\cdots,l-1\}$
\State \quad \quad (2) Compute $ x_{i+1,j_0} = h \left(N|h(M)|H_i\cdot z^T_{i,j_0} \oplus x_{i,j_0}\right)$ for $i\in\{0,1,\cdots,l-1\}$
\State \quad \quad (3) Accept if $ x_{l,j_0} = x_{0,j_0}$ and reject otherwise.
\end{algorithmic}
}
\end{algorithm}

\textbf{Security and Efficiency.} The ZLC scheme is based on CFS signatures, whose security relies on two assumptions: It is hard to solve an instance of the SD problem, and it is hard to distinguish a Goppa code from a random one -- the GD problem. The authors of \cite{ZhengLC07} also showed that the ZLC construction provides unforgeability and anonymity. Indeed, the probability of forging a signature is $\frac{1}{2^n}$, and any adversary outside the ring cannot guess the signer's identity due to the uniform distribution of $x_{i,j_0}$. This scheme is as efficient as the CFS signature, and verification takes $tl$ column operations\footnote{One column operation is one access to a table plus one operation like a comparison or an addition}
 and $l+1$ hash computations; the total signature length is close to $(n-k)+ \log_2({n \choose t})l$ bits, where $\log_2({n \choose t})$ is the number of bits required to address a word of length $n$ and weight $t$. For instance, for $m = 16$ and $t = 9$, the signature length is about $144 + 126l$ bits.

\subsection{Threshold Ring Signatures}

Since its introduction in 2001, a lot of effort has gone into modifying and extending the ring signature scheme \cite{RST2001}. One such extension is the BSS threshold ring signature scheme first proposed by Bresson, Stern and Szydlo \cite{BSS02} in 2002. In threshold ring signature schemes, the secret signing key is distributed amongst $N$ members; at least $l$ of these members are required to generate a valid signature. More precisely, in an $(l,N)$ threshold signature scheme, any set of $l$ members can generate an $l$-out-of-$N$ signature on behalf of the whole group, without revealing their identity. This type of construction decreases the cost of signing, as it does not require the participation of all $N$ members.

Several threshold ring signatures have followed \cite{BSS02}. For example, Wong et al. \cite{WongFLW03} proposed the tandem construction, a threshold signature scheme using a secure multiparty trapdoor transformation. The threshold ring signature in \cite{LiuWW03} uses both RSA- and DL-based public keys at the same time and introduces the notion of separability: all signers can select their own keys independently, with distinguishable parameter domains.
These signatures, however, and many others, are factoring- ECC-, or pairing-based. Only two coding-based proposals are known, however, up to date. In the following, we outline these proposals.

\subsubsection{Aguilar et al.'s scheme (ACG)} The first non-generic code-based threshold ring signature scheme is introduced in \cite{MelchorCG08}; it generalizes Stern's identification protocol into a threshold ring signature scheme, using the Fiat-Shamir paradigm \cite{FiaSham86}. Algorithm \ref{ACGTRSS} explains how Aguilar et al.'s construction works. We denote by $N$ the number of signers (provers) in the ring, and let $l$ with $(l\leq N)$ stand for the number of first signers. A leader $\mathrm{S_L}$ amongst them gives to the ring members their public keys.

\begin{algorithm}
\caption{The ACG Identification Scheme }\label{ACGTRSS}
\begin{algorithmic}
\State \textbf{Key Generation:} Each potential signer $S_i$ has:
\State \quad \quad- \textbf{The public key:} the $ (n-k)\times n$ binary matrix $H_i$
\State \quad \quad- \textbf{The private key:} $n$-bit word $s_i$ the weight $t$ s.t. $H_is_i^T =0$,
\State \quad \quad- \textbf{The ring public key :} $ (n-k)N\times nN $ binary matrix $H$ defined by:
\begin{center}
$
H =
\begin{pmatrix}
  H_1 & 0 & \cdots & 0\\
  0 & H_2 & 0 & 0\\

  \vdots& \ddots&H_i &0\\
  0&0 &\cdots&H_N
\end{pmatrix}
$
\end{center}
\State \textbf{Commitment:}
\State \quad \quad - Each prover $S_i$ (among $l$ signers) chooses randomly $z_i\in \FF{2}^n$ and a permutation $\sigma_i$ of $\{1,\cdots,n\}$
\State \quad \quad - Each prover $S_i$ sends to $S_L$ three commitments $c_{1,i} $, $c_{2,i}$ and $c_{3,i}$ given by:
\State \quad \quad \quad \quad $c_{1,i} = h(\left( \sigma_i, H_iz_i^T \right))$, $c_{2,i} = h(\sigma_i(z_i))$ and $c_{3,i} = h(\sigma_i \left( z_i \oplus s_i\right))$

\State \quad \quad  - $S_L$ generates the $N-l$ missing commitments for the $N-l$ non-signers by fixing all remaining $s_i$ at $0$.

\State \quad \quad - $S_L$ chooses randomly a constant $n$-block permutation $\Pi$ on $N$ blocks

\State \quad \quad - $S_L$ computes the master commitments $C_1$, $C_2$ and $C_3$ using $c_{1,i}$, $c_{2,i}$ and $c_{3,i}$ by:

\State \quad \quad \quad \quad $C_1 = h(\Pi \left(c_{1,1},\cdots,c_{1,N}\right))$, $C_2= h(\Pi \left(c_{2,1},\cdots,c_{2,N}\right))$,
$C_3= h(\Pi \left(c_{3,1},\cdots,c_{3,N})\right)$

\State \quad \quad - $S_L$ sends $C_1$, $C_2$ and $C_3$ to the verifier $V$.

\State \textbf{Challenge:}
\State \quad \quad -$V$ sends a challenge $b\in\{0,1,2\}$ to $S_L$ which forwards this challenge to $l$ signers.

\State \textbf{Response:}
\State \quad \quad - Perform the challenge step of the Stern's protocol between each prover $S_i$ and $S_L$
\State \quad \quad - $S_L$ simulates the missing $N-l$ Stern's protocol with $s_i=0$ for all $l+1\leq i\leq N$
\State \quad \quad - $S_L$ gathers all answers to create the global response for $V$ as follows:
\State \quad \quad \quad \quad \quad * If $b=0$: $S_L$ sets $z=(z_1\cdots,z_N)$, $\Omega = \Pi \circ (\sigma_1,\cdots,\sigma_N)$ and reveals $z$ and $\Omega$
\State \quad \quad \quad \quad \quad * If $b=1$: $S_L$ constructs $ x = (y_1 \oplus s_1,\cdots,y_N \oplus s_N)$ and reveals $x$ and $\Omega$
\State \quad \quad \quad \quad \quad * If $b=2$: $S_L$ constructs $\Pi(y_1, \cdots,y_N)$ and reveals $\Omega(s_1, \cdots,s_N)$

\State \textbf{Verification:}
\State \quad \quad - If $b=0$: $V$ checks that $\Omega(s)$ is a $n$-block permutation and that $C_1$, $C_2$ were honestly computed.
\State \quad \quad - If $b=1$:  $V$ checks that $\Omega(s)$ is a $n$-block permutation and that $C_1$, $C_3$ were honestly computed.
\State \quad \quad - If $b=2$:  $V$ checks that:
\State \quad \quad \quad \quad \quad * $C_2$, $C_3$ were honestly computed
\State \quad \quad \quad \quad \quad * $\wt(\Omega(s)) = lt$
\State \quad \quad \quad \quad \quad * each of block of $\Omega(s)$ of length $n$ has weigth $t$ or $0$.
\end{algorithmic}
\end{algorithm}

\textbf{Security and Efficiency.} Aguilar et al.'s identification scheme is a zero-knowledge protocol with a cheating probability of $2/3$ as in Stern's scheme. Its security relies on the hardness of the SD problem: finding a vector $s \in \mathbb{F}_2^{nN}$ of weight $tl$ and a null syndrome w.r.t. $H$ such that each block (out of $N$) of length is of weight $t$ or $0$. The signing complexity and signature length are $N$ times those of Stern's signature scheme: a complexity of about $140n^2N$ independently of $l$, and a length of about $20kB\times N$. In order to reduce the public key size, \cite{MelchorCG08} suggested the use of double-circulant matrices, requiring $nN/2$, rather than $n^2N/2$ storage bits. For double-circulant matrices, \cite{gaborit-girault} proposes parameters $n = 347$ and $t = 76$ for an $83$-bit security level, rather than $n = 634$, $t = 69$, and a rate $1/2$, as in an $80$-bit secure Stern's scheme.

\subsubsection{Dallot et al.'s scheme } A second code-based threshold ring signature has been proposed by Dallot and Vergnaud (DV) in \cite{dallot-vergnaud}, combining the generic construction of Bresson et al. \cite{BSS02} with the CFS signature scheme. The DV construction requires the following: an $(n,k)$ $t$-error-correcting binary Goppa code with $n = 2^m$ and $k = n-mt$, where $m$ a positive integer. We denote by $N$ and $l$ the number of ring users and the number of signers respectively. Let $h$ be a public collision-resistant hash function of range $\left\{0,1\right\}^{mt}$, $f_{(\cdot)}$, a trapdoor one-way function : $\left\{0,1\right\}^{a}\rightarrow \left\{0,1\right\}^{mt}$, and $(E_{k,i})$, a family of random permutations that encrypts $b$-bit messages with $a_{0}$-bit keys and an additional parameter $i\in$ $[1, N]$. We again denote concatenation as $(s|s')$ and random selection by $ x \xleftarrow[]{R} \mathcal{S}$.

For simplicity, we index the signers as $1,\cdots, l$. In addition, each ring member $i$ is associated with a secret/public key pair as in the CFS construction, i.e. the public key is $H_i = Q_i\widetilde{H_i}P_i$ and the secret key is $(P_i,\widetilde{H_i},Q_i,\gamma_i)$. Dallot et al.'s procedure is presented in Algorithm \ref{TRSDV}.

\begin{algorithm}
\caption{The DV threshold ring signature scheme}\label{TRSDV}
\begin{algorithmic}
\State \textbf{Key Generation :} Each signer in the ring has to:
\State \quad - choose an $(n,k)$-code $\mathcal{C}_i$ over $\mathbb{F}_2$ having a decoding algorithm $\gamma_i$ correcting up to $t$ errors.
\State \quad - construct an $n\times (n-k) $ parity check matrix $\widetilde{H_i}$ of $\mathcal{C}_i$.
\State \quad - choose randomly an $(n-k)\times(n-k)$ invertible matrix $Q_i$ over $\mathbb{F}_2$.
\State \quad - choose randomly an $n\times n$ permutation matrix $P_i$ over $\mathbb{F}_2$.
\State \quad \quad - \textbf{The public key:} $H_i = Q_i\widetilde{H_i}P_i$
\State \quad \quad - \textbf{The private key:} $(P_i,\widetilde{H_i},Q_i,\gamma_i)$
\State \textbf{Signature:} To generate a signature on a message $M$:
\State \quad - compute the symmetric key for $E$: $k=h(M)$.
\State \quad - compute value at origin: $v_0 =h(H_i,\cdots,H_N)$ .
\State \quad - choose random seeds: For each $i=l+1,\cdots,N$ do
\State \quad \quad \quad (1) $ x_{i} \xleftarrow[]{R} \{x\in \mathbb{F}_2^{n}\quad \text{s.t.} \quad \wt(x)\leq t\}$
\State \quad \quad \quad (2) $ r_{i} \xleftarrow[]{R} \{1\cdots,2^{tm}\}$
\State \quad \quad \quad (3) $ y_i \xleftarrow[]{} H_i x_{i}^T + h(M|r_i)$
\State \quad - compute a sharing polynomial: Find a polynomial $f$ over $\mathbb{F}_{2^{tm}}$ s.t.
\State \quad \quad \quad - $\deg(f)=N-l$
\State \quad \quad \quad - $f(0)=v_0$
\State \quad \quad \quad - $f(i)=E_{k,i}(y_i)$ \,\,$\forall\,l+1 \leq i\leq N$
\State \quad - For each $i=1,\cdots,l$ do
\State \quad \quad \quad - $ x_{i} \xleftarrow[]{} \emptyset$
\State \quad \quad \quad - While $ x_{i} = \emptyset$ do
\State \quad \quad \quad \quad (1) $ r_{i} \xleftarrow[]{R} \{1\cdots,2^{tm}\}$
\State \quad \quad \quad \quad (2) $ z_{i} \xleftarrow[]{} \gamma_{i}(Q_i^{-1}\cdot(E_{k,i}(f(i))+ h(M|r_i)))$
\State \quad \quad \quad \quad (3) if $ z_{i}\neq \emptyset$ then $x_{i} \xleftarrow[]{} z_iP_i^{-1}$
\State \quad - \textbf{The signature:} $\sigma=(N,x_{1}, \ldots,x_{N},r_{1}, \ldots,r_{n},f)$
\State \textbf{Verification:} Given $(N,x_{1}, \ldots,x_{n},r_{1}, \ldots,r_{n},f)$ any user can verify the signature by:
\State \quad - Recovering the symmetric key: $k=h(M)$
\State \quad - Recovering $(y_i)$: $y_i =  H_i x_{i}^T + h(M|r_i)$
\State \quad - checking the equations:
\State \quad \quad \quad (1) $f(0)\stackrel{\mathrm{?}} = h(H_1,\cdots,H_N)$
\State \quad \quad \quad (2) $f(i) \stackrel{\mathrm{?}} = E_{k,i}(y_{i})$ \,\,$\forall\,1 \leq i\leq N$
\end{algorithmic}
\end{algorithm}

\textbf{Security and Efficiency.} The DV construction is a provably secure threshold ring signature satisfying three properties: consistency, anonymity, and unforgeability \cite{dallot-vergnaud}. Unforgeability is proved based on two coding theory problems. One is the well known NP-complete \cite{berlekamp-mceliece-vantilborg} Bounded Distance Decoding problem (GBDP) which is a variant of the SD problem with the constraint that the number of errors is up to $(n-k/\log_2(n))$ as in the mCFS signature scheme. The second is the GCD problem: distinguishing a randomly sampled Goppa code from a random linear code (with the same parameters); this problem is widely considered as difficult \cite{Sen02}. The complete security proof of the DV scheme is in \cite{dallot-vergnaud}.\\

For a ring with $N$ members, the set of public keys $(H_i)$ are stored in $n(n-k)N$ bits. To produce a valid signature, the signer has to perform the following calculations: computing $N-l$ syndromes, $N$ polynomial evaluations that can be performed in $2N(N-l)$ binary operations using Horner's rule and $l(t!)$ decodings of Goppa codes, each consisting of: computing a syndrome (in about $t^2m^2/2$ binary operations), computing a localisator polynomial ($6t^2m$ binary operations) and computing its roots ( $2t^2m^2$ binary operations). Thus, the total cost for generating a signature would be $(N-l)t^2m^2/2 + 2N(N-l)+ l(t!)(3/2 + 6/m)$ binary operations.

Signature verification requires $(N+1)$ polynomial evaluations $N$ syndrome-computations, resulting in $2(N+1)(N-l) + Nt^2m^2/2$ binary operations. The signature consists of: the number $N$ of ring-users, which are stored in $\log_2(N)$ bits, $N$ random vectors $x_i$ of weight up to $t$ which can be indexed with a $\lfloor \log_2\sum_{i=1}^{t} 	{2^m \choose i}\rfloor$ bit counter, $N$ random vectors $r_i$ in $\{0,\ldots,2^{mt}-1\}$ requiring at most $mt$ bits and a polynomial of degree $N-l$ which needs $ (N-l+1)mt$ bits. The signature size is thus about $N\left(\lfloor \log_2\sum_{i=1}^{t} 	{2^m \choose i}\rfloor + 2mt\right) + \log_2(N) -(l-1)mt$ bits.

\subsection{Blind signatures}

Blind signatures were first introduced by Chaum \cite{Chaum82} for applications such as e-Voting or electronic payment systems, which require anonymity. The main goal of Chaum's scheme is to ensure \textit{Blindness} (i.e., the signed message is disguised -- blinded -- before signing) and \textit{Untraceability} (i.e., the signer cannot trace the signed message after the sender has revealed the signature publicly).

Several blind signature schemes followed Chaum's proposal. In 1988, the authors of \cite{ChaumFN88} showed a new signature scheme for electronic payment systems. Later, the authors of \cite{StPiCa95} introduced fair blind signature schemes. In 1992, another blind signature scheme based on factoring and discrete logarithm-based identification schemes \cite{Oka92} have been developed. Based on Schnorr's \cite{Schnorr91} and Guillou-Quisquater's \cite{GuiQui88} protocols, provably secure blind signature schemes were presented in \cite{PS96}.
As far as we know, there exists only a single code-based blind signature scheme, namely Overbeck's construction \cite{RO09}.

\subsubsection{Overbeck's scheme} The general idea behind Overbeck's protocol is, instead of blinding the message, to use permuting kernels in order to blind the signer's public key from a public key of a code. A blind signature is thus generated by the owner of a valid secret key, with the blinded public key. During verification, the blinder gives a static zero-knowledge proof showing that the private and public keys are paired. This proof is based on the Permuted Kernels Problem (PKP) which can be formulated as follows: Given a random $(n,k)$ code and a random permuted subcode of dimension $L < k$, find the permutation. This problem is known to be NP-hard in the general case \cite{ShamirPKP89}.

For simplicity, we denote a code by its generator matrix. Let $h$ be a hash function, $r$ be a random seed, and $w$ be a positive integer. Denote by PKP-proof$(A,B)$ the static PKP-Proof that code $A$ is an isometric subcode of code $B$, s.t. $dim(A)\leq dim(B)$. The notation $dim(C)$ stands for the dimension of the code $C$.

A slightly modified version of Overbeck's blind signature scheme is depicted in Algorithm \ref{OverbeckBlind}.

\begin{algorithm}
\caption{Overbeck's blind signature}\label{OverbeckBlind}
\begin{algorithmic}
\State \textbf{Key Generation:}
\State \quad - choose an $(n,k)$-code $\mathcal{C}$ over $\mathbb{F}_2$ having a decoding algorithm $\gamma$ correcting up to $t$ errors.
\State \quad - construct an $(n-k)\times n $ parity check matrix $\widetilde{H}$ of $\mathcal{C}$.
\State \quad - choose randomly an $(n-k)\times(n-k)$ invertible matrix $Q$ over $\mathbb{F}_q$.
\State \quad - choose randomly an $n\times n$ permutation matrix $P$ over $\mathbb{F}_q$.
\State \quad \quad - \textbf{The public key:} $H = Q\widetilde{H}P$
\State \quad \quad - \textbf{The private key:} $(P,\widetilde{H},Q,\gamma)$
\State \textbf{Blinding:} The user has to:
\State \quad - generate a random $p\times n$ public matrix $R_0$ over $\mathbb{F}_q$
\State \quad - generate a random $L\times p$ matrix $K$ of full rank over $\mathbb{F}_q$
\State \quad - set $R =KR_0$
\State \quad - generate a $n \times n$ permutation matrix $\Pi$
\State \quad - create the blind generator matrix $G_b$ as follows: $G_b = [\frac{G}{R}]\Pi$
\State \quad - derive from $G_b$ the blind check matrix $H_b$
\State \quad - solve $H_bx^T=h(M|H_b)$ in $x$
\State \quad - output $s = H(x\Pi^{-1})$ (the blind syndrome) and $u = (r,\Pi,H_b)$ (the unblinding information)
\State \textbf{Unblinding:} Given $M$, $s$ and a correct signature $\sigma$ of $H$:
\State \quad - Check if that: $\wt(\sigma)= t $ and $H\sigma^T = s$. If not output \texttt{failure}.
\State \quad - Generate a PKP-Proof$(H_b,H_0)$ with $H_0 = [\frac{H}{R_0}]$
\State \quad - Output the blind signature $\sigma_b = (r,H_b,\sigma \pi,\text{PKP-Proof}(H_b,H_0))$
\State \textbf{Verification:} Given $r$, $M$, and $\sigma_b = (r,H_b,\sigma \pi,\text{PKP-Proof}(H_b,H_R))$, where $H_R$ is a parity-check matrix of the code generated by $R$, verify $\sigma_b$ by:
\State \quad - Generate the matrix $R_0$ from $r$
\State \quad - Find some vector $\tau$ satisfying $H \tau^T = h(M|H_b)$
\State \quad - Verify $\wt(\sigma \Pi)< t$ and $(\tau-\sigma \Pi)\in H_b$
\State \quad - Check PKP-Proof$(H_b,H_R)$
\end{algorithmic}
\end{algorithm}

\textbf{Security and Efficiency.} Overbeck assessed the efficiency of his scheme by applying it to the CFS construction. For a $(2^m,2^m-mt)$ binary Goppa code, the complexity of this scheme is as follows: To store a public parity check matrix key $H$ $2^m\times mt$ bits are needed. The blind matrix $H_b$ is a $2^m\times (mt-L)$ binary matrix. To generate a single signature, the Blinding algorithm is run $\left(2^{mt}/{2^m \choose t}\right)$ times, each time requiring $m^3t^2$ binary operations for the signer and $m^3t^3$ for the blinder. Thus, the total signing complexity is about $\left(2^{mt}/{2^m \choose t}\right)(m^3t^2 + m^3t^3)$ binary operations per signature. The blind signature size mainly depends on PKP-Proof$(H_b,H_0)$ requiring the storage of the generator matrix $G_b$ of size $((k+L) \times n)$ bits in each round.

The author of \cite{RO09} does not explicitly prove the proposed construction, but he claims that the scheme is provably secure based on the hardness of some instances of the PKP and SD problems.

\subsection{Identity-based Signatures}

Identity-based cryptography was proposed by Shamir in 1984 \cite{SHA84} so as to simplify PKI requirements. An identity is associated with data such as an e-mail or IP-address instead of a public key; the secret key is issued by a trusted Key Generation Center (KGC) thanks to a master secret that only the KGC knows. Some PKI and certificate costs can now be avoided. However, identity-based cryptography suffers from a major drawback: the KGC must be trusted completely. A solution to this problem, also known as the key escrow problem,  is to employ multiple PKGs that jointly produce the master secret key (see \cite{Bo01}).

Identity-based cryptography has led to the development, in 1984, of identity-based signature (IBS) schemes. One of the most interesting contributions to this subject is the framework of \cite{BNG04}, for which a large family of IBS are proved secure. This work was later extended in \cite{GalindoHK06}, which implied the existence of generic IBS constructions with various additional properties that are provably secure in the standard model.
\subsubsection{Cayrel et al.'s identification scheme} The first coding-based IBS appeared in \cite{cayrel-gaborit-girault} is due to Cayrel, Gaborit and Girault (CGG). The main idea of this scheme is to combine the mCFS scheme with a slightly modified Stern scheme to obtain an IBS scheme whose security relies on the syndrome decoding problem. The mCFS scheme is used to solve an instance of the SD problem given a hash value of an identity in the first step, while Stern's protocol is used for identification in the following step.\\

Consider a linear $(n,k)$-code over $\mathbb{F}_2$ with a disguised parity-check matrix $H$ defined by $H = Q\widetilde{H}S$ with $\widetilde{H}$ the original parity-check matrix, $Q$ invertible, and $S$ a permutation matrix. The matrix $H$ is public, while $Q$ and $S$ are kept secret by a trusted Key Generation Center (KGC). Denote by $h$ a hash function with outputs in $\{0,1\}^{n-k}$. In addition, let $y$ be the identity associated to the prover wishing to authenticate to a verifier. The Cayrel et al. identification scheme works as Stern's protocol, with a few variations.

\textbf{Security and Efficiency.} Cayrel et al.'s identification scheme (CFS-Stern IBS) is provably secure against passive (i.e., eavesdropping only) impersonation attacks \cite{cayrel-gaborit-girault}, based on the hardness of the SD and GD problems. The security and the performance of the proposed identification scheme mainly depends on the difficulty of finding a couple $\{s,j\}$ without the description $H$. At the same time, an attacker needs to minimize the number of attempts used to find $j$, so as to be able to find $s$ with minimal cost.

\begin{algorithm}
\caption{The CGG Identity-based scheme} \label{CGG}
\begin{algorithmic}
\State \textbf{Key Deliverance:}
\State \quad - The prover sends its identity $y$ to KGC
\State \quad - TCA runs the CFS algorithm (Alg. \ref{CFSAlgo}) on $y$ to get $\{s,j\}$ s.t. $h(h(y)|j) = Hs^T$ with $\wt(s)\leq t$
\State \quad \quad- \textbf{The Public key:} $h(h(y)|j)$
\State \quad \quad- \textbf{The Private key:} $\{s,j\}$
\State \textbf{Identification:} Run the Stern's protocol as follows:
\State \quad - \textbf{Commitments:}
\State \quad \quad \quad - $P$ chooses randomly $u \text{ from } \mathbb{F}_2^n$ and $\sigma$ permutation over $\left\{ 1, \ldots, n\right\}$
\State \quad \quad \quad - $P$ computes the commitments $c_1$, $c_2$ and $c_3$ as follows:
\State \quad \quad \quad \quad \quad \quad $c_1 = h\left( \sigma, Hu^T \right)$, $c_2 = h(\sigma(u))$, $c_3 = h(\sigma \left( u \oplus s\right))$
\State \quad \quad \quad - $P$ sends $c_1$, $c_2$ and $c_3$ and $j$ to $V$
\State \quad - \textbf{Challenge:} $V$ choose randomdy $b\in\{0,1,2\}$ and sends it to $P$
\State \quad - \textbf{Response:}
\State \quad \quad \quad - If $b=0$: $P$ sends $u$ and $\sigma$ to $V$
\State \quad \quad \quad - If $b=1$: $P$ sends $u \oplus s$ and $\sigma$ to $V$
\State \quad \quad \quad - If $b=2$: $P$ sends $\sigma(u)$ and $\sigma(s)$ to $V$
\State \quad - \textbf{Verification :}
\State \quad \quad \quad - If $b=0$: $V$ checks if $c_1$ and $c_2$ were honestly computed
\State \quad \quad \quad - If $b=1$: $V$ checks if $c_1$ and $c_3$ were honestly computed
\State \quad \quad \quad - If $b=2$: $V$ checks if $c_3$ and $c_3$ were honestly computed and $\wt(s) =t$
\end{algorithmic}
\end{algorithm}
\subsection{Summary}
In Table \ref{tab:classic} we summarize the complexity of the code-based proposals with special properties by using the following notations: $t$ is the correction capability of the code, $n$ denotes the code length which equals $2^m$ in the case of Goppa codes, $k$ indicates the code dimension, $N$ is number of users in the ring, $l $ is number of involved signers in the ring , $L$ is dimension of the subcode introduced in \cite{RO09} and $r_i$ is number of rounds for $i =1,2$.

Based on these notations, we define the two following quantities:
\begin{itemize}
	\item $A(m,t,N,l) = N\left(\lfloor \log_2\sum_{i=1}^{t}{2^m \choose i}\rfloor + 2mt\right) + \log_2(N) -(l-1)mt$
 	\item $B(m,t,N,l)= (N-l)t^2m^2/2 + 2N(N-l)+ l(t!)(3/2 + 6/m)$.
\end{itemize}

\begin{table}
\centering
	\caption{Code-based signatures with special properties with parameters $(m,t,N,l,L,r_1,r_2) = (15, 12, 100, 50, 40, 58, 80)$}
	\label{tab:classic}
	\scalebox{0.85}{
    \begin{tabular}{ | l |l | l | l | }
    \hline
   \textbf{Schemes}& \textbf{Pk size in bits}& \textbf{Sign. size in bits} & \textbf{Sign. cost in bops} \\ \hline  \hline
    \textbf{Identity Based Signatures }& & &  \\ \hline
PGGG \cite{PGGG0}  & $2^mtm$ ($\approx$ 0.7 MB) & $2^m\times r_1$ ($\approx$ 1.1 MB)& $ t!t^2m^2(1/2 + 2 + 6/m)$($\approx 2^{45}$)\\ \hline
\textbf{Ring Signatures }& & &  \\ \hline
ZLC \cite{ZhengLC07}& $2^mtm$ ($\approx$ 0.7 MB) & $tm+ \log_2({2^m \choose t})l$ ($\approx$ 0.95 kB)& $t!t^2m^2$ ($\approx 2^{43.8}$)  \\ \hline
\textbf{Threshold(ring) Signatures }& & &  \\ \hline
ACG \cite{MelchorCG08}& $n^2N/2 $ ($\approx$ 2.41 MB) & $20 000\times N$ ($\approx$ 0.24 MB) & $140n^2N$ ($\approx 2^{32.3}$)   \\ \hline
DV \cite{dallot-vergnaud}&  $2^mtm N$ ($\approx$ 70 MB)  & $A(m,t,N,l)$ ($\approx$ 5.2 kB) & $B(m,t,N,l)$ $ (\approx 2^{35.4}$)\\ \hline
\textbf{Blind Signatures }& & &   \\\hline
Overbeck\cite{RO09}& $2^mtm$ ($\approx$ 0.7 MB)  & $((2^m-tm+L) 2^m)\times r_2 $ ($\approx$ 9.95 GB) & $\left(2^{mt}/{2^m \choose t}\right)(m^3t^2 + m^3t^3)$ ($\approx 2^{190}$)\\\hline
\end{tabular}
}
\end{table}

\newpage
\section{Conclusion}\label{sec:conclusion}
Several code-based signature schemes already exist, exhibiting features such as small public key size (Stern \cite{stern}), short signature size (CFS \cite{courtois-finiasz-sendrier}), or a good balance of public key and signature size at the expense of security (KKS \cite{kks97}). By combining such schemes, additional constructions such as identity-based, threshold ring, or blind signatures can be obtained. However these schemes also inherit the disadvantages of the underlying protocols. We strongly encourage the code-based research community to actively investigate future possibilities for post-quantum signature schemes, such as multi-signatures, group signatures, or linkable signatures.

\bibliographystyle{plain}
\bibliography{Code-based_Bib}

\end{document}